\documentclass[12pt]{article}
\usepackage{graphicx}
\begin{document}
\oddsidemargin 0in
\evensidemargin 0in
\topmargin 0in
\centerline{\bf AB INITIO CALCULATIONS OF RESPONSE PROPERTIES}
\vskip 0.05cm
\centerline{\bf INCLUDING THE ELECTRON-HOLE INTERACTION} 
\par
\vskip 1cm
\noindent
VALERIO OLEVANO$^{*,**}$,STEFAN ALBRECHT$^{*}$, LUCIA REINING$^{*}$
\par\noindent
$^{*}$ Laboratoire des Solides Irradi\'es,
   UMR 7642 CNRS -- CEA,
   \'Ecole Polytechnique, F-91128 Palaiseau, France 
\par\noindent
$^{**}$ Istituto Nazionale per la Fisica della Materia, Dipartimento di
Fisica
   dell'Universit\`a \\ di Roma ``Tor Vergata'',
   Via della Ricerca Scientifica, I--00133 Roma,
   Italy 
\par\vskip 0.5truecm \noindent


\par\noindent
{\bf ABSTRACT}
\par\vskip 0.5truecm \noindent
We discuss the current status of a computational approach which 
allows to evaluate the dielectric matrix, and hence electronic excitations
like optical properties, including local field and excitonic effects. 
We introduce a recent numerical development which greatly reduces the
use of memory in such type of calculations, and hence eliminates
one of the bottlenecks for the application to complex systems. 
We present recent applications of the method, focusing our interest
on insulating oxides.   

\par \vskip 0.5truecm\noindent
{\bf INTRODUCTION}
\par \vskip 0.5truecm\noindent
Density-functional theory (DFT) in the Local Density Approximation
(LDA) is widely and successfully used as a state-of-the-art tool
to compute the ground-state electronic properties of many-electron
systems \cite{hohk}.  It is feasible to apply DFT to
systems as complex as surfaces, defects, and clusters.
However, while the ground-state properties can in principle be obtained
exactly within DFT, many spectroscopic properties are in general not directly
accessible in such a calculation.

In fact, it is well known that the use of the DFT-LDA eigenvalues as the
physical energies  entering in the absorption process relies on several
approximations. First of all, Kohn-Sham LDA eigenvalues are only a rough
approximation of the electron addition or removal energies, such those
measured in photoemission. The correct energies should be obtained by using
the true electron self--energy operator $\Sigma$, which appears at
the place of the DFT--LDA exchange--correlation potential in an equation
similar to the Kohn-Sham one. A good approximation for $\Sigma$, which allows
to determine the quasiparticle  (QP) bandstructure, can be obtained within 
Hedin's GW approach \cite{hedin}. In this scheme, DFT-LDA results can be used as
the starting point, in a first-order perturbative approach 
\cite{hyblou,godshsh}.

However, the true QP energies, together with the LDA wavefunctions,
 are in principle still not sufficient
to describe correctly an absorption process, in which electron-hole pairs
are created. Their interaction can lead to bound exciton states
which occur within the gap, and  can also induce appreciable distortions of the
spectral lineshape above the continuous--absorption edge. 

Since recently \cite{na4,AlbOnRein_Li20_97,noi,SHIRLEY_1,SHIRLEY_2,LOUIE},
these excitonic effects can be 
treated in the {\it ab initio} framework.
Various applications have demonstrated the success of the Green's
functions approach 
\cite{na4,AlbOnRein_Li20_97,noi,SHIRLEY_1,SHIRLEY_2,LOUIE,SHIRLEY_CAF2,examples}. 
However, the calculations are still
very cumbersome, since a two-particle problem has to be solved. This prevents
the method from being applied on a large scale.

In this work, we introduce a perturbative approach to the solution
of the Bethe-Salpeter equation which greatly reduces the
use of memory in such type of calculations, and hence eliminates
one of the bottlenecks for the application to complex systems.
We show that the approximation is perfectly controllable, hence the approach is
still {\it ab initio}. 
In order to test the limits of the method, 
we present recent applications for the case 
of insulating oxides.
\par
\newpage

\noindent{\bf THE APPROACH}
\par\vskip 0.5truecm\noindent  
The absorption spectrum is given by the imaginary part of the macroscopic
dielectric function $\epsilon_M$
\begin{equation}
\epsilon_M(\omega ) = 1 - \lim_{{\bf q} \to 0} v({\bf q}) \hat \chi_{{\bf G}%
=0,{\bf G}^{\prime}=0} ({\bf q};\omega ),
\end{equation}
where $\hat \chi ({\bf r},{\bf r}^{\prime};\omega ) = -iS({\bf r},{\bf r},%
{\bf r}^{\prime},{\bf r}^{\prime};\omega )$.
The four-point function $S$ obeys the
Bethe-Salpeter equation, 
\begin{equation}
S(1,1^{\prime };2,2^{\prime })=S_{0}(1,1^{\prime };2,2^{\prime
})+S_{0}(1,1^{\prime };3,3^{\prime })\Xi (3,3^{\prime };4,4^{\prime
})S(4,4^{\prime };2,2^{\prime }).
\end{equation}
 The notation (1,2) stands
for two pairs of space and time coordinates, $({\bf r}_1, t_1;{\bf r}_2,
t_2) $.
Repeated arguments are integrated over. The term $S_{0}(1,1^{\prime
};2,2^{\prime })=G(1^{\prime },2^{\prime })G(2,1)$ yields the polarization
function of independent quasiparticles $\chi _{0}$, from which the standard
RPA $\epsilon _{M}$ is obtained
($G(1,1^{\prime})$ is the one-particle Green's function \cite{HS}).
The kernel $\Xi $ contains two
contributions:
\begin{equation}
\Xi (1,1^{\prime },2,2^{\prime })=-i\delta (1,1^{\prime })\delta
(2,2^{\prime })v(1,2)+i\delta (1,2)\delta (1^{\prime },2^{\prime
})W(1,1^{\prime }).
\end{equation}

Considering the first term in the calculation of $S$ is equivalent to the
inclusion of local field effects in the matrix inversion of a standard RPA
calculation. In order to obtain the macroscopic dielectric constant, the
bare Coulomb interaction $v$ contained in this term must, however, be used
without the long range term of vanishing wave vector \cite
{hankedelsolefiorino84}. When spin is not explicitly treated, $v$ gets a
factor of two for singlet excitons. In the second term, $W$ is the screened
Coulomb attraction between electron and hole. It is obtained as a functional
derivative of the self-energy in the $GW$ approximation, neglecting a term $G%
\frac{\delta W}{\delta G}$. This latter term contains information about the
change in screening due to the excitation, and is expected to be small \cite
{strinati}. We limit ourselves to static screening, since dynamical effects in the
electron--hole screening and in the one particle Green's function tend to
cancel each other \cite{bech2}, which suggests to neglect both of them.
The set of equations (1) - (3) are at the basis of all the
{\it ab initio} exciton calculations which have appeared in the literature
recently 
\cite{na4,AlbOnRein_Li20_97,noi,SHIRLEY_1,SHIRLEY_2,LOUIE,SHIRLEY_CAF2,examples}. 

In order to solve Eq.\ (2), we 
rewrite it as an effective eigenvalue problem, 
\begin{equation}
\sum_{(n_3,n_4)}H_{exc}^{(n_1,n_2),(n_3,n_4)} 
A_{\lambda}^{(n_3,n_4)} = E_{\lambda } A_{\lambda}^{(n_1,n_2)},
\end{equation}
with 
\begin{eqnarray}
H_{exc}^{(n_{1},n_{2}),(n_{3},n_{4})} &=&(E_{n_{2}}-E_{n_{1}})\delta
_{n_{1},n_{3}}\delta _{n_{2},n_{4}}-i(f_{n_{2}}-f_{n_{1}})\times  \nonumber
\\
&&\int \cdot \int d{\bf r}_{1}\, d{\bf r}_{1}^{\prime }\, d{\bf r}_{2}\, d{\bf r}%
_{2}^{\prime }\ \psi _{n_{1}}({\bf r}_{1})\,\psi _{n_{2}}^{*}({\bf r}%
_{1}^{\prime })\,\Xi ({\bf r}_{1},{\bf r}_{1}^{\prime },{\bf r}_{2},{\bf r}%
_{2}^{\prime })\,\psi _{n_{3}}^{*}({\bf r}_{2})\,\psi _{n_{4}}({\bf r}%
_{2}^{\prime }).
\end{eqnarray}
The $\psi _{n}({\bf r})$ are LDA Bloch functions, with $n$ denoting a
band index and a Bloch vector ${\bf k}$. For the calculation
of absorption spectra, we can limit ourselves to transitions with positive
frequency, i.e. $(n_{1},n_{2})$ and $(n_{3},n_{4})$ are pairs of one valence
and one conduction band, respectively (in other words, we consider only
the resonant part). Moreover, we build up the spectra of optical properties by
considering only negligible momentum transfer, hence the same ${\bf k}$
for the valence and the conduction state. 
Equation (1) reads then:

\begin{equation}
\epsilon _{M}(\omega )=1+\lim_{{\bf q}\to 0}v({\bf q})\sum_{\lambda }\frac{%
\left| \sum_{v,c;{\bf k}}
\langle v,{\bf k}-{\bf q}|e^{-i{\bf qr}}|c,{\bf k}\rangle
A_{\lambda
}^{(v,c;{\bf k})} \right|^{2}}{(E_{\lambda }-\omega )}.
\end{equation}

\newpage\noindent
\underline{\bf The problem}
\par\vskip 0.5truecm\noindent
The advantage of this approach with respect to a straightforward
inversion of equation (2) is that we have to diagonalize the Hamiltonian (5)
only once, and can then construct the spectrum for all frequencies.
Moreover, the knowledge of the coefficients $A_{\lambda}^{(v,c;{\bf k})}$
allows for a detailed analysis
of the results. However, the matrix to be diagonalized can be very large,
since the basis set is built up of {\it pairs} of states. Moreover, 
the number of k-points in the Brillouin zone sampling needed for the 
calculation of optical properties is typically at least one order of
magnitude bigger than that employed in a ground state calculation.
Even though this number enters the basis linearly, and not quadratically
as the number of bands, 
the basis is very large,
and the calculations are prohibitive for an application to an extended range
of energies, i.e. including states far 
from the Fermi level, or to systems with many
atoms.

In order to find a way for overcoming this difficulty, it is worthwhile
to analyse how excitonic effects alter the absorption spectra.

We first look at bulk Silicon. The importance of excitonic effects in
the absorption spectrum of Silicon have already been shown by Hanke
and Sham in a semi-empirical calculation \cite{HS}. Recently,
{\it ab initio} calculations \cite{noi,SHIRLEY_1}
have confirmed their findings.  For this work, we have repeated the
calculations of Ref. \cite{noi}, but using a k-point set for the
Brillouin zone (BZ) sampling which is shifted
off the high symmetry directions, whereas in Ref. \cite{noi} special points
\cite{MonPak76}
were used. It has in fact been discussed \cite{COMMENT} that the shifted grid
gives improved results. Still, the number of k-points which must be used
in order to get reasonable results is big; here we have chosen 256 points
in the BZ, which, together with the 4 valence
bands and the 3 lowest conduction bands, leads to a matrix size of
3072$\times$3072. (It is obvious that increasing the size of the unit cell
or the energy range which is considered would let the problem explode
rapidly.) 
\begin{figure}
\includegraphics[width=12cm]{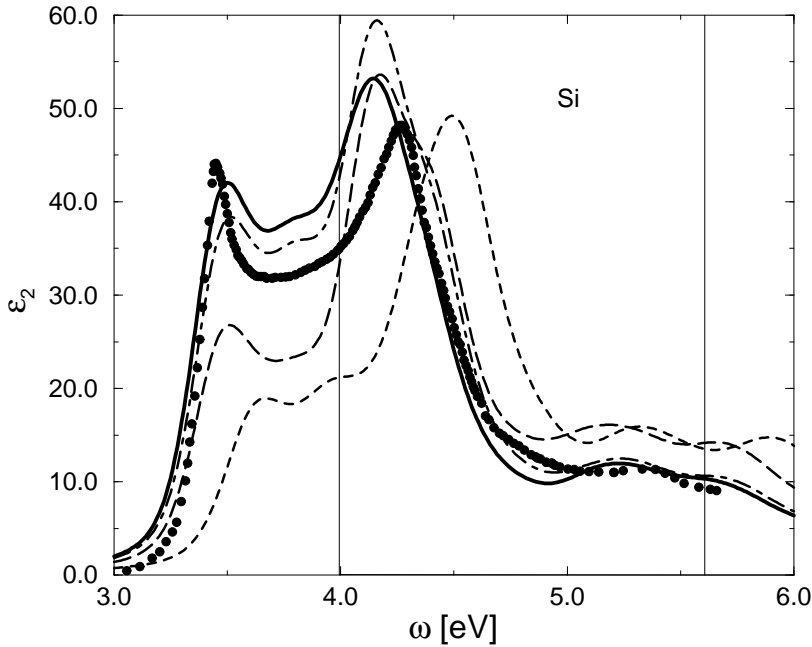}
\caption{Absorption spectrum of bulk silicon. Dots: Experiment \cite{Laute87}. 
Short-dashed curve: calculation, no electron-hole interaction included. 
Continuous curve: full exciton calculation. Long-dashed curve: 
zero-order result. Dot-dashed curve: first order result. Vertical 
bars: block cuts. }
\label{SIEPS}
\end{figure}
\par\vskip 0.5truecm\noindent
\underline{\bf Analysis of the excitonic effect}
\par\vskip 0.5truecm\noindent
The continuous curve in Fig. \ref{SIEPS} shows hence the absorption spectrum of 
bulk silicon
which we obtain using the parameters described above.
It is in good agreement with experimental data \cite{Laute87} (dots in Fig. 1), 
and with the previous
{\it ab initio} calculations \cite{noi,SHIRLEY_1}. We can now analyse 
the origin of the excitonic effects. 
First of all, the fact that excitonic effects change the peak positions
in Silicon can be misleading: in fact, the density of transition energies
does virtually not change between the independent quasiparticle and 
the exciton result. This has been found not only for silicon, but 
seems to hold quite generally for those materials where no strongly
bound excitons exist (see, for example, \cite{SHIRLEY_CAF2,examples}). 
The effect is hence entirely due to a change in oscillator
strength, in other words, to the coefficients $A_{\lambda}^{(vc)}$
from equation (4). $A_{\lambda}^{(vc)}$ expresses the mixing of
different independent quasiparticle transitions with energy $E_c - E_v$ 
into a given transition with energy $E_{\lambda}$. 
\begin{figure}
\includegraphics[width=12cm]{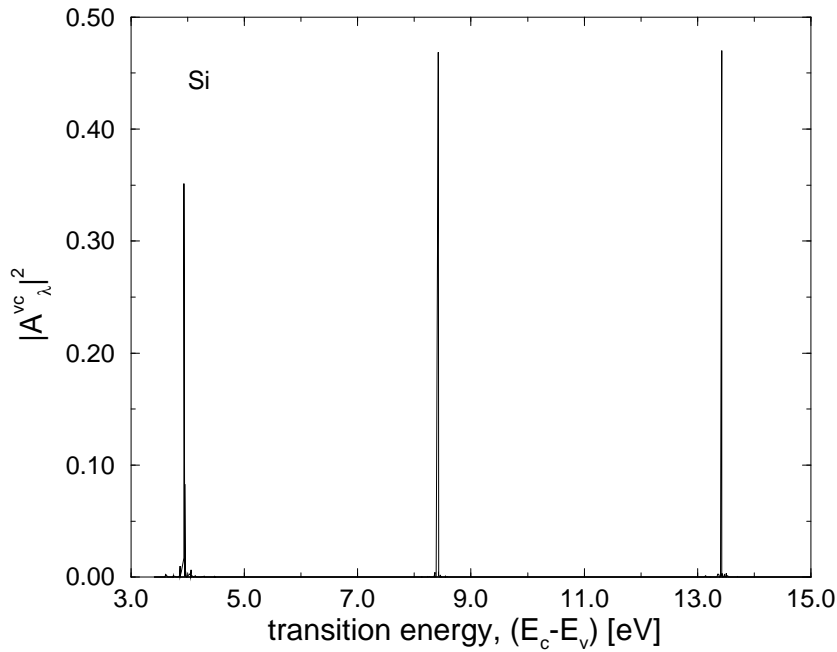}
\caption{Contribution of independent-quasiparticle transitions to 
excitonic transitions at given energy (see text), for bulk silicon}
\label{SIEX}
\end{figure}
Fig. 2 shows the
absolute squared value $\vert A_{\lambda}^{(vc)}\vert ^2$ for three
different $\lambda$, with $E_{\lambda}$ chosen above the independent 
quasiparticle gap by
0.5 eV, 5 eV and 10 eV, respectively, in function of the energy of the
contributing independent-quasiparticle transitions, $E_c - E_v$. 
One can notice that the distributions
are extremely sharp; in fact more than 90 \% of the weight is contained in
an energy range around each peak position of less than
0.1 eV, and more than 99 \% of the weight is contained in
a region of less than 1 eV. 
This suggests at first sight that one should divide the spectrum into 
several parts, and for each energy range carry out the calculations using 
the corresponding limited part of the hamiltonian only.  
\par\vskip 0.5truecm\noindent
\underline{\bf The perturbative approach}
\par\vskip 0.5truecm\noindent
In other words, one could hope that a separate diagonalisation 
of several subblocks, neglecting the interaction between the
blocks, could yield a satisfactory absorption spectrum for all energies
except those very close to the boarder of the blocks. This is however
not true: In Fig. 1 
 we compare the result obtained by exact diagonalisation of the full
3072$\times$3072 Hamiltonian (continuous curve) with that  
obtained by choosing about 40 blocks in a reasonable way,
i.e. by avoiding to cut at energies close to maxima in the
density of states. The result
is given by the long-dashed curve. The vertical lines indicate the
block cuts. The agreement of the result obtained in this way
with the full calculation is not 
satisfactory at any energy; in particular, when we compare with the 
result obtained by neglecting the electron-hole interaction 
(short-dashed curve) we notice that a big part
of the excitonic effect has been lost.
 In order to obtain the full effect, it is hence 
unavoidable to let the states interact even at relatively large energy
distances. 
We can however assume that the interaction decreases with increasing
energy separation, and try to include the missing contributions in a
perturbative approach, as also suggested in Ref.\cite{PROC} . 

In fact, we can treat that portion of the excitonic
hamiltonian, which has been neglected up to now by our choice of the blocks,
in first order perturbation theory for quasi-continuum states. In this
way, we obtain 
the dot-dashed curve in Fig. 1: the agreement with the full result is now
very good.
Note that in this way, we have to diagonalize matrices of a maximum size
of 100$\times$100, whereas the full calculation requires
the diagonalization of a 3072$\times$3072 matrix. This represents a memory
saving of about three orders of magnitude, which will be very useful for the
study of bigger systems, or in order to be able to improve  the
k-point sampling. 
It must be stressed that the quality of the resulting
spectrum can be checked and systematically
improved by increasing the block size of
the zero-order calculation. 
 
\par\vskip 0.5truecm\noindent
{\bf RESULTS FOR INSULATING OXIDES}
\par\vskip 0.5truecm\noindent
This new approach looks hence extremely promising, and it can be expected that 
the method will work similarly well for systems with the same behaviour
as silicon: 
silicon has in fact the peculiarity that 
screening is strong, hence the electron-hole interaction matrix elements
$W_{vc}^{v'c'}$ are weak. 
The strong deformation of the spectrum comes from the fact that many 
transitions at energies very close to each other, where
\par\noindent 
$W_{vc}^{v'c'}/[(E_c-E_v) - (E_{c'}-E_{v'})]$ is big, are interacting.
This point is of course very favourable for our
approach, and it is interesting to explore whether the method can also be
applied in a less favourable situation.
In order to become more severe, we can choose an insulator, with low 
dielectric constant,
and in the following we will hence look at two insulating oxides: MgO and
Li$_2$O. 
\par\vskip 0.5truecm\noindent
\underline{\bf MgO}
\par\vskip 0.5truecm\noindent
Excitonic effects in the absorption spectrum of magnesium oxide have been
calculated {\it ab initio} by Shirley et al \cite{SHIRLEY_2}, using the 
Haydock recursion method \cite{Haydock80} for a direct solution of
equation (2).
Excitonic effects have turned out to alter the absorption
spectrum significantly. 
With a dielectric constant of only about $\epsilon_M = 3$,
MgO is hence a good candidate for our test. 
We have performed the ground state calculations using pseudopotentials
of the Trouiller-Martins type \cite{MartTrou91}, a plane-wave cutoff of 50 Ry,
and using 10 special points in the irreducible part of the
BZ. The theoretical lattice constant is
4.13 A in good agreement with the experimental value of 4.21 A. We have then 
determined the GW corrections to the DFT-LDA eigenvalues, using 307
reciprocal lattice vectors and 150 bands. The correction to the direct
gap turns out to be 2.5 eV, and the corrections are slightly larger at
the zone boundaries, in good agreement with \cite{ARY} and \cite{SHIRLEY_GW}.
\begin{figure}
\includegraphics[width=12cm]{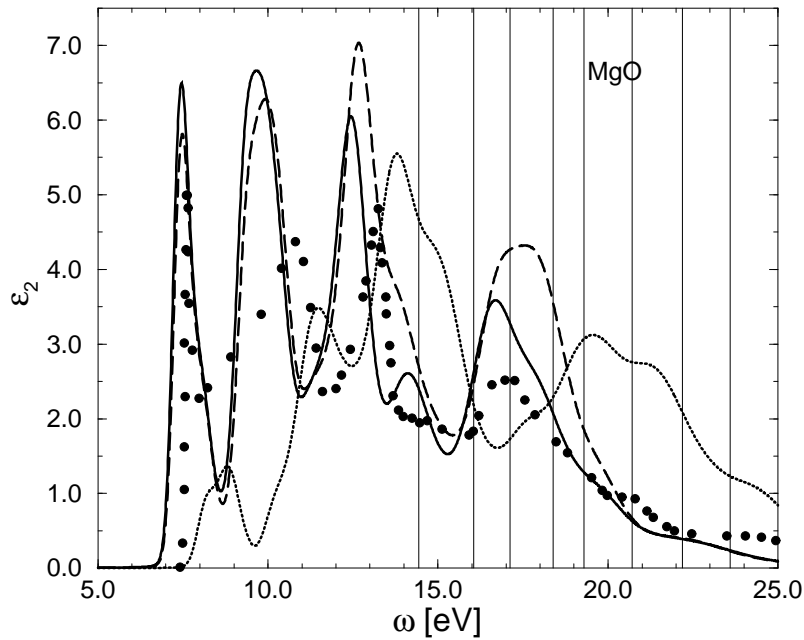}
\caption{Absorption spectrum of magnesium oxide. Dots: Experiment 
\cite{MGO_EXP}. 
Dotted curve: calculation, no electron-hole interaction included. 
Continuous curve: full exciton calculation. Dashed curve: 
 first order result. Vertical 
bars: block cuts.}
\label{MGOEPS}
\end{figure}
Our independent quasi-particle absorption
spectrum, calculated using 256 shifted k-points in the BZ and the first 9
bands,
(dotted curve in Fig. 3) is close to the result
of Ref. \cite{SHIRLEY_2}, and in very bad agreement with
experiment (dots in Fig. 3) \cite{MGO_EXP}. We have then calculated the  
full exciton absorption spectrum according to equation (6), using
the same parameters as for the independent quasiparticle spectrum. 
The result 
is given by the 
continuous curve in Fig. 3, and is again close to the findings of 
Ref. \cite{SHIRLEY_2}.
As in that work, the very strong 
excitonic effect significantly improves the agreement
with  experiment \cite{MGO_EXP,FOOT_1}. The main discrepancy is the 
overestimation of the peak intensities, which might be due to our 
limited k-point sampling.  
Now we want to try our perturbative approach.
\begin{figure}
\includegraphics[width=12cm]{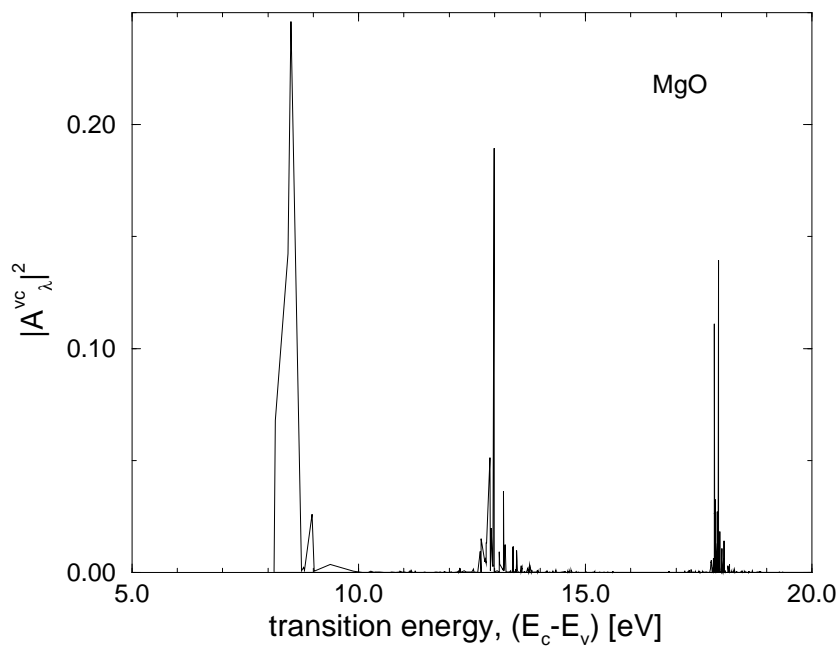}
\caption{Contribution of independent-quasiparticle transitions to 
excitonic transitions at given energy (see text), for magnesium oxide}
\label{MGOEX}
\end{figure} 
First, we can check if this might be reasonable, by analysing
the exciton eigenstates as we did in the case of silicon.
This analysis is shown in Fig. \ref{MGOEX}. 
Also in MgO, the distributions are looking rather sharp, 
but of course, due to the
stronger $W_{vc}^{v'c'}$,  
transitions from
a wider range of the spectrum must be taken into account.
This time, about 90 \% of the
weight of the peaks is contained in a range of  less than 1.5 eV,
and in order to retain 99 \% of the weight we must go up to almost 7 eV
for the second peak (3 eV for the first one and 2 eV for the third one). 
This tells us that we certainly have to choose bigger blocks
than in Silicon in order to obtain reasonable results. However, still
we will try to reduce the memory
requirement by about two orders of magnitude, and therefore we choose the
block borders as indicated by the vertical lines in Fig. 5. This corresponds
to blocks of a maximum size of 500$\times$500. 
The resulting first-order spectrum is given by the
dashed curve. Again, the agreement with  
the full calculation is good, and most of the disagreement
between the independent quasiparticle and the experimental result has been
removed. 
\begin{figure}
\includegraphics[width=12cm]{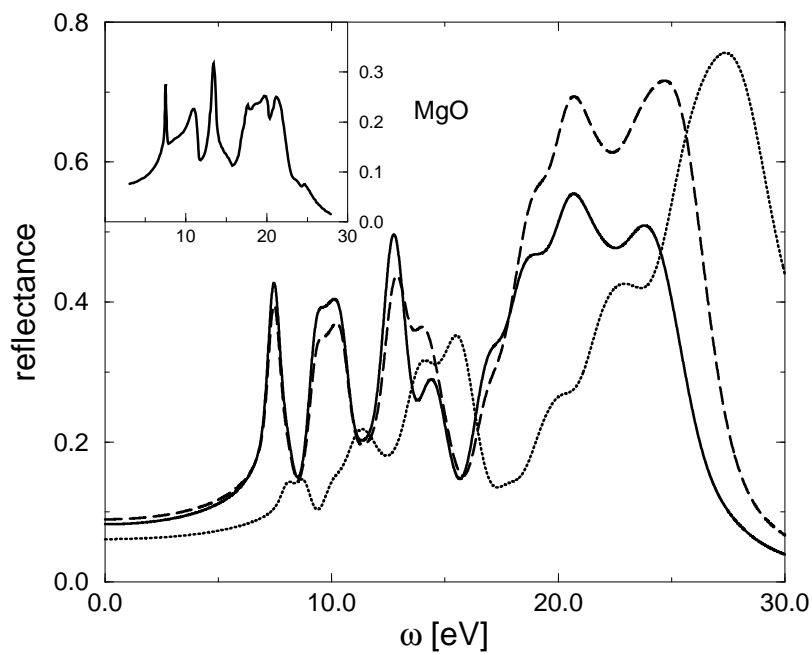}
\caption{Reflectance spectrum of magnesium oxide. Inset: Experiment 
\cite{MGO_EXP}. 
Dotted curve: calculation, no electron-hole interaction included. 
Continuous curve: full exciton calculation. Dashed curve: 
 first order result.}
\label{MGOREF}
\end{figure} 
The most important errors occur of course at higher energies, where the
states are denser and hence the energy range of the blocks smaller.
Due to the Kramers Kronig relation, one could suspect that this fact
might spoil the real part of $\epsilon_M$ even at low energies, and hence
show up in spectra linked to both the real and the imaginary part,
like reflectance. We have therefore used our results in order to
calculate the reflectance spectrum of MgO. The independent quasiparticle
result is given by the dotted curve in Fig. \ref{MGOREF}, and the 
result of the full exciton calculation is given by the continuous curve. 
As in the case of $Im(\epsilon_M)$, the inclusion of the electron-hole
interaction has a dramatic effect. In particular, the weak structures
up to 15 eV in the GW spectrum are transformed into pronounced
peaks, which are consistent with the experimentally
found structures in that energy range in the reflectance spectrum 
(see inset) \cite{MGO_EXP}.
The first-order result, for the block size of 500$\times$500, is given
by the dashed curve. Again, the agreement with the
full calculation  at lower energies is excellent, and also the structures
at higher energies are considerably improved with respect to the GW
calculation. The performance of the method for MgO
is hence less spectacular than in the case of silicon, but still
the approach turns out to be valid.

\par\vskip 0.5cm\noindent
\underline{\bf Li$_2$O}
\par\vskip 0.5cm\noindent
Another insulator of interest for our purpose is lithium oxide, Li$_2$O.
Li$_2$O is a material which is experimentally studied essentially because
of its potential importance in the atomic energy domain. Several spectroscopic
measurements are published in the literature\cite{RADI}, 
often focusing on the 
characterisation of radiation defects. Recently, there have been
theoretical \cite{AlbOnRein_Li20_97} and experimental \cite{ISHII} 
studies devoted to the question of excitonic effects at the band edge
of Li$_2$O. In Ref. \cite{AlbOnRein_Li20_97} a lowering of the optical
gap due to excitonic effects of the order of 1 eV was predicted. The experiment
of Ref. \cite{ISHII} seems to be in agreement with this finding. It is hence 
interesting, on one side, to study excitonic effects in Li$_2$O not
only at the absorption onset, but over a 20 eV range, and to compare with
experiment. On the other hand, the fact that now also the transition energies
are heavily affected by the electron-hole interaction presents an additional
test for our approach.
\begin{figure}
\includegraphics[width=12cm]{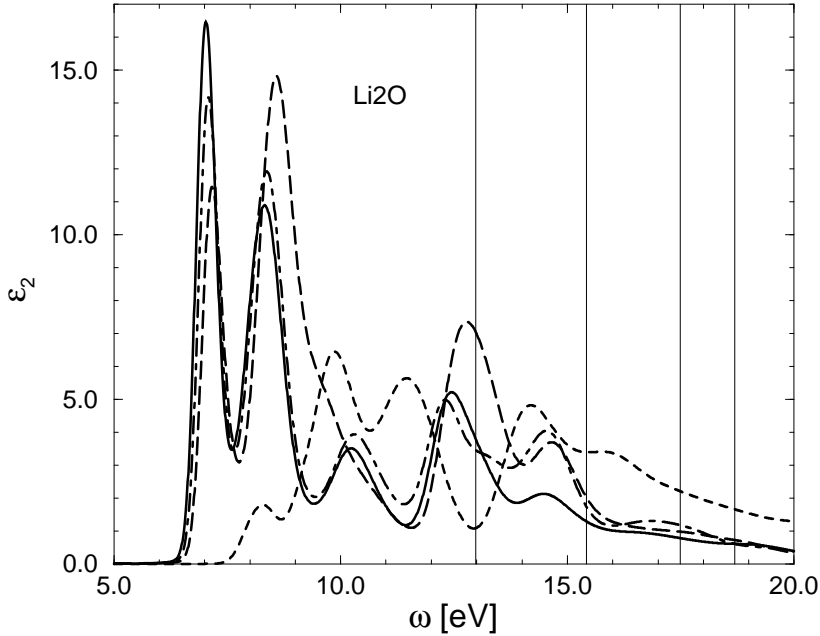}
\caption{Absorption spectrum of lithium oxide.
Short-dashed curve: calculation, no electron-hole interaction included. 
Continuous curve: full exciton calculation. Long-dashed curve: 
 first order result, block size 300. Dot-dashes curve: first 
 order result, block size 700. Vertical 
bars: block cuts}
\label{LI2OEPS}
\end{figure}
We have performed the ground state calculations using pseudopotentials
of the Trouiller-Martins type \cite{MartTrou91}, a plane-wave cutoff of 80 Ry,
and using 2 special points in the irreducible part of the
BZ. The theoretical lattice constant is
4.534 A in good agreement with the experimental value of 4.573 A. We have then
determined the GW corrections to the DFT-LDA eigenvalues, using up to
541 reciprocal lattice vectors and 280 bands. The correction to the direct
LDA gap of 5.6 eV turns out to be 2.3 eV, again with corrections which 
are slightly larger at
the zone boundaries, in good agreement with Ref. \cite{AlbOnRein_Li20_97}.

Our independent quasiparticle absorption
spectrum, calculated using 256 shifted k-points in the BZ and the first 9
bands,
(short-dashed curve in Fig. \ref{LI2OEPS}) consequently
shows an onset at about 7.9 eV, which well compares with the 
value of 7.99 eV estimated in Ref. \cite{ISHII} for the band-gap energy,
and is considerably higher than the measured onset at about 6.5 - 7 eV
\cite{RAUCH,ISHII,RADI}.
We have then calculated the
full exciton absorption spectrum according to equation (6), using
the same parameters as for the independent quasiparticle spectrum.
The result
is given by the
continuous curve in Fig. \ref{LI2OEPS}. The onset has now been lowered, to
about 7 eV, which agrees well with the experiment and with the prediction of
Ref. \cite{AlbOnRein_Li20_97}. Moreover, as in the case of silicon and MgO,
considerable oscillator strength has been transferred to lower energies, 
giving rise to a sharp peak at the onset. This behaviour is similar to what has
been found for LiF in Ref. \cite{SHIRLEY_2}.
\begin{figure}
\includegraphics[width=12cm]{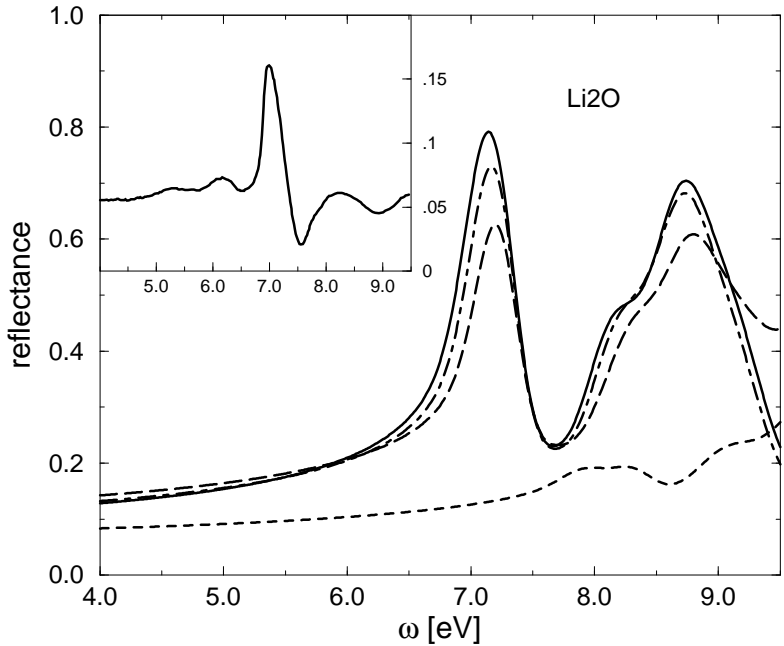}
\caption{Reflectance spectrum of lithium oxide. Inset: experiment \cite{ISHII}.
Short-dashed curve: calculation, no electron-hole interaction included.   
Continuous curve: full exciton calculation.
Long-dashed curve: first order result, block size 300. Dot-dashed curve: 
first order result, block size 700.}
\label{LI2OREF}
\end{figure}
Our calculated reflectance spectrum, given by the continuous curve in Fig.
\ref{LI2OREF}, reproduces the main features of the experimental result
\cite{ISHII} (see inset Fig. \ref{LI2OREF}),
 with a first strong peak at 7 eV, a second broader peak
above 8 eV and a sharp minimum between the two structures. A detailed 
analysis will be necessary in order to check whether the small structure on the
low energy shoulder of the 8 eV-peak is related to the shoulder found 
experimentally, and whether the excitonic origin of this structure, 
proposed in Ref. \cite{ISHII},  can
be confirmed. 

For this work, we concentrate on the question whether our perturbative
approach can also be applied to this system with large excitonic binding
energy.  
We have therefore also calculated the first order spectrum, given by 
the long-dashed curve  in Fig. \ref{LI2OEPS}. Here the block size 
has been chosen to be 300$\times$300. The essential part of the 
excitonic effect is already included in this spectrum. In
particular, the binding energy of the first peak is given correctly.
 As mentioned 
above, we can improve this result systematically by increasing the 
block size: The dot-dashed curve has been calculated with a block size 
of 700$\times$700. The vertical bars are the corresponding block cuts. 
The result is now in very good agreement with the full one.
The same holds for the reflectance spectrum, as can be seen from the 
short-dashed (GW), long-dashed (300$\times$300) and dot-dashed (700$\times$700) 
curves in Fig.\ref{LI2OREF}.
This means that, even in a very unfavourable situation, our approach 
is advantageous. 
\par\vskip 0.5truecm\noindent
{\bf OUTLOOK}
\par\vskip 0.5truecm\noindent
We should stress the fact that up to now we have concentrated our  effort
only at the aim to save memory. It is however obvious that we can in principle
use the same method also to save considerable CPU time and disk storage 
requirements: the biggest part of the exciton Hamiltonian $H^{exc}$ is now not
longer used in a direct diagonalization, but, for the calculation
of $A_{\lambda}$, in the perturbation formula
\par\noindent
$\sum_t H^{exc}_{t\lambda }/(E_{\lambda} - E_t) A_t^{(0)}$, with 
$H^{exc}_{tt'} = \sum_{(vc),(v'c')} A_t^{*(0)(vc)} H_{(vc)}^{exc,(v'c')}
A_{t'}^{(0)(v'c')}$. The fact that now $H$ acts on a vector, instead of
being inverted or diagonalized, allows to change space for each part of
the Hamiltonian separately.
In particular, one can go back to real space for the electron-hole attraction
part $W$, which, according to Eq. (3), is then diagonal in two of the four 
coordinates. In this way, our approach will hence combine both the advantages
of the effective two-particle picture used in \cite{na4,noi,LOUIE}, and
of the recursive inversion approach proposed in \cite{SHIRLEY_1,SHIRLEY_2}. 

\par\vskip 0.5truecm\noindent {\bf CONCLUSIONS}
\par\vskip 0.5truecm\noindent
In conclusion, we have analysed in detail the nature of the excitonic
effects in different materials. This has allowed us to design 
a method which improves the efficiency of
the {\it ab initio}
calculation of optical spectra including the electron-hole interaction.
By using a perturbative approach, the method avoids the diagonalisation of
large matrices and therefore has the advantage of significant memory saving.
Still, it is {\it ab initio} in the sense that one can converge
the results smoothly to the exact ones, by a controllable parameter (i.e.
the block size).

The method works extremely well for systems with weak interaction, but
strong excitonic effects, as we have shown for the case of bulk silicon.
We can therefore expect that it will be of great usefulness for the
study of systems like those, for example, built up of many silicon atoms
(e.g. amorphous silicon), where we are in the ideal condition of weak
interaction but large density of transitions.

We have tested the approach also for unfavourable cases, using as examples
two insulators. In spite of the strongly increased interaction strength, we
can still obtain good spectra in the gap region, with memory savings
of two orders of magnitude. We have also shown that the method yields
valid results for the real part, not only the imaginary part, of $\epsilon_M$,
therefore allowing for the description of spectra like the reflectance.
The inclusion of first order corrections to a given subset of states
is hence a very promising method in order to obtain, with little effort,
improved spectra in the low energy region. 
\par\vskip 0.5truecm\noindent
{\bf ACKNOWLEDGEMENTS}
\par\vskip 0.5truecm\noindent
For this work, computer time was granted by IDRIS (Project No. CP9/990544).


\begin{thebibliography}{99}
\bibitem{FOOT_1}  In order to discuss the remaining
discrepancies, we would have to converge the spectrum better, especially
what the BZ sampling is concerned. This is however beyond the scope
of this work.
\end{thebibliography}
\end{document}